\documentclass[12pt, twocolumn]{openjournal}
\usepackage{graphicx,float}
\usepackage[dvipsnames]{xcolor}
\usepackage{hyperref}
\hypersetup{
    colorlinks=true,
    linkcolor=blue,
    filecolor=blue,      
    urlcolor=blue,
    citecolor=blue,
}
\usepackage{color,colortbl}
\DeclareGraphicsExtensions{.bmp,.png,.jpg,.pdf}
\usepackage{orcidlink}
\usepackage{amsmath}
\usepackage{framed}
\begin{document}
\title{Photon (Non)Conservation in the Reduced Speed of Light Approximation and How to (Almost) Fix It}
\author{Nickolay Y. Gnedin\orcidlink{0000-0001-5925-4580}}
\affiliation{Theory Division; 
Fermi National Accelerator Laboratory;
Batavia, IL 60510, USA}
\affiliation{Kavli Institute for Cosmological Physics;
The University of Chicago;
Chicago, IL 60637, USA}
\affiliation{Department of Astronomy \& Astrophysics; 
The University of Chicago; 
Chicago, IL 60637, USA}

\begin{abstract}
The "Reduced Speed of Light" (RSL) approximation is commonly used to speed up radiative transfer calculations in cosmological simulations. However, it has been shown previously that the RSL approximation leads to photon non-conservation when the radiation field is rapidly evolving in time. I show that these missing photons can be counted exactly for some numerical schemes. Adding them back into a simulation, however, is a much harder task. I show one example of such a scheme, which achieves sub-percent accuracy on simple tests. Unfortunately, the scheme performs much worse on semi-realistic simulations of cosmic reionization, leading to a faster overlap and significant errors in the point-wise comparison of the RSL radiation field with the reference simulation that maintains the full speed of light for the radiative transfer.
\end{abstract}


\maketitle

\section{Motivation}

The past decade has witnessed a dramatic expansion in our observational window onto the early Universe. The James Webb Space Telescope (JWST) has revealed a rich population of galaxies within the first few hundred million years after the Big Bang, while wide-field surveys have uncovered an increasing number of luminous quasars at redshifts beyond 7. Together, these discoveries are transforming our empirical understanding of the reionization era and providing stringent new constraints on when and how the intergalactic medium was ionized.

At the same time, cosmological simulations have achieved remarkable progress in reproducing many aspects of galaxy formation and the large-scale structure of the Universe. Yet, despite these advances, our understanding of cosmic reionization remains limited by the accuracy of radiative transfer methods. As reionization is fundamentally a radiative process, even small numerical errors in photon transport can propagate into large uncertainties in key observables—from the global ionized-fraction history to the morphology of ionized regions and the mean free path of ionizing photons.

This is particularly apparent in comparing two of the recent "second-generation" simulations, "CROC" \citep{Gnedin2014,GnedinKaurov2014}, and "THESAN" \cite{Kannan2022,Garaldi2022,Garaldi2024}. The two sets of simulations are similar in many respects (box sizes, spatial and mass resolution, modeled physical processes, numerical methods, etc). The primary difference in modeled physics between the two is the specific implementation of star formation and stellar feedback. Nevertheless, and completely unexpectedly, the power spectra of the radiation source luminosity density are similar in two simulations, while the power spectra of the actual radiation field are surprisingly different \citep{Gnedin2025}. Moreover, the power spectra of the radiation field in CROC and THESAN can be matched on large scales if one allows the timing in two simulations to be selected independently - i.e., for each CROC snapshot, one can find a snapshot from THESAN at a different cosmic time with a similar large-scale power spectrum of the radiation field. 

It appears, therefore, that in two simulations, radiation sources are clustered similarly and emit photons at similar rates, but these photons propagate at different rates through the modeled IGM. When phrased in such a way, it is actually not surprising, as both simulations (a) use a "reduced speed of light" (RSL) approximation \citep{Gnedin2001,Gnedin2014,Kannan2019} in modeling radiative transfer, and (b) employ different approximate numerical methods for solving the radiative transfer. Either of the two differences could be the culprit for the confusing behavior of the radiation field in two simulations (the lack of similarity at the same time and similarity on large scales at different times). The goal of this paper is to explore further the photon conservation (or the lack thereof) in the RSL approximation, while exploring the effects of different numerical schemes is a more complex endeavour and is left for future work. Some of the tests in this paper indeed utilize different numerical schemes, but that exploration is not comprehensive and is not essential for the conclusions of this work.

This work builds on and extends the prior work by \citet{Cain2024}, who, to the best of my knowledge, was the first to explicitly show that the RSL approximation does not conserve photons in some limits. Even earlier, the effect was mentioned by \citet{Ocvirk2019}, although they did not interpret it as a lack of photon conservation.

\section{RSL Primer}

The radiative transfer equation for the (monochromatic) radiation specific intensity $I_\nu(\vec{x},\vec{n},t)$ in a static background reads
\begin{equation}
    \frac{1}{c}\frac{\partial}{\partial t}I_\nu + \frac{\partial}{\partial x^j} n^i I_\nu = 
    -\kappa_\nu I_\nu + \frac{\dot{E}_\nu}{4\pi},
    \label{eq:rt}
\end{equation}
where 
\[
    \kappa_\nu = \sum_j \sigma_j(\nu) n_j
\]
is the absorption coefficient per unit length, the sum is over all atomic species that absorb radiation at frequency $\nu$, and $E_\nu$ is the monochromatic luminosity density of radiative sources. With the cosmological expansion, additional terms appear on the left-hand side, which I ignore for now, since the deliberations below apply equally to a static background and the cosmologically expanding one. These cosmological terms can always be considered as just parts of a more complex time derivative operator.

For monochromatic radiation, the difference between the number density and the energy density of photons is a trivial rescaling; therefore, I will switch to the photon number density below, as it may be more familiar to the cosmological radiative transfer community.

The first three moments of $I_\nu$ are
\begin{subequations}%
\label{eq:momdef}%
\begin{align}
    N_\nu(\vec{x},t) & = \frac{1}{ch\nu} \int d\Omega\, I_\nu(\vec{x},\vec{n},t), \\
    M_\nu^i(\vec{x},t) & = \frac{1}{ch\nu} \int d\Omega\, I_\nu(\vec{x},\vec{n},t) n^i = \langle n^i\rangle N_\nu, \\
    P_\nu^{ij}(\vec{x},t) & = \frac{1}{ch\nu} \int d\Omega\, I_\nu(\vec{x},\vec{n},t) n^i n^j = \langle n^i n^j\rangle N_\nu. 
\end{align}
\end{subequations}
The photon number flux $F^i \equiv c M^i$. I will, however, use the first\footnote{Being a computational physicist, my counting starts at 0.} moment $M^i$ instead of the flux since it will be important to explicitly consider all factors of $c$.

Hereafter, I will drop the frequency subscript for clearer notation, as all radiation moments will be monochromatic unless the frequency dependence becomes explicitly relevant (as in the calculation of the photo-ionization rate).

In this notation, the equations for the first 2 moments become:
\begin{subequations}%
\label{eq:moms}%
\begin{align}
    \frac{\partial N}{\partial t}+ c\frac{\partial M^j}{\partial x^j} & = 
    -c \kappa N + \dot{N}, \\
    \frac{\partial M^i}{\partial t} + c\frac{\partial P^{ij}}{\partial x^j} & = 
    -c \kappa M^i + \dot{M}^i, 
\end{align}
\end{subequations}
where $\dot{N} \equiv \dot{E}/(h\nu)$ is the photon production rate density and $c\dot{M}^i$ is the flux source, which vanishes for isotropic sources - the assumption I adopt hereafter (note that hereafter the dot over a symbol is not a time derivative but a part of the notation).

In the Newtonian limit ($c\rightarrow \infty)$, the radiation field can be computed in a closed form - just sum up $1/(4\pi R^2)$ from each source,
\begin{equation}
    N_{\rm NL}(\vec{x}) = \frac{1}{4\pi c} \int d^3 y \frac{\dot{N}(\vec{y})}{(\vec{x}-\vec{y})^2} e^{-\tau(\vec{x},\vec{y})},
    \label{eq:exa}
\end{equation}
where
\[
    \tau(\vec{x},\vec{y}) = |\vec{x}-\vec{y}|\int_0^1 \kappa\left(\alpha\vec{x}+(1-\alpha)\vec{y}\right) d\alpha
\]
is the optical depth between locations $\vec{x}$ and $\vec{y}$.

The well-known issue with equations (\ref{eq:moms}) is that photons propagate at the speed of light, and hence any explicit numerical scheme must have the time-step smaller than the light crossing time of one resolution element (a cell on the grid for a grid code). This time-step is going to be too small for the vast majority of realistic applications. Hence, the "Reduced Speed of Light" (or RSL) approximation is used to circumvent this numerical limitation.

In order to demonstrate explicitly how the RSL is introduced, Equation (\ref{eq:moms}a) can be written in an equivalent form, where I explicitly label two places where the speed of light enters:
\begin{equation}%
\label{eq:mom0}%
    \frac{1}{c_a}\frac{\partial N}{\partial t} + \frac{\partial M^j}{\partial x^j} = 
    - \kappa N + \frac{\dot{N}}{c_b}.
\end{equation}
In the current literature, there appear to exist two different flavors of the RSL approximation:
\begin{description}
\item[A] $c_a = \hat{c}$, $c_b=c$ \citep{Gnedin2001,Gnedin2014},
\item[B] $c_a = c_b = \hat{c}$ \citep{Rosdahl2013,Kannan2019},
\end{description}
where $\hat{c} \ll c$ is the reduced speed of light.

Sometimes it is also useful to consider the spatially averaged version of Equation (\ref{eq:moms}a),
\begin{equation}%
\label{eq:avg}%
    \frac{1}{c_a}\frac{d}{dt} \langle N\rangle = 
    - \bar{\kappa} \langle N\rangle + \frac{\langle\dot{N}\rangle}{c_b},
\end{equation}
where \emph{by definition}
\[
\bar\kappa \equiv \frac{\langle\kappa N\rangle}{\langle N\rangle}.
\]

In the following, I assume that $\hat{c}$ is either constant in space or a sufficiently weak function of spatial position (like jumping only on refinement boundaries) that one can move $\hat{c}$ in and out of spatial derivatives without worrying about gradients of $\hat{c}$. If this is not the case, the derivations presented below will need to be modified appropriately.

Both flavors of the RSL approximation have their pros and cons.

\smallskip
{\parindent=0pt \bf Flavor A:} 
\begin{leftbar}
\begin{equation}%
\label{eq:moma}%
    \frac{\partial N_A}{\hat{c}\partial t} + \frac{\partial M_A^j }{\partial x^j} = 
    - \kappa N_A + \frac{\dot{N}}{c}.
\end{equation}
\begin{description}
\item[Pros] In a steady-state $(\partial N_A/(\hat{c}\partial t) \rightarrow 0)$,  $N_A = N_{NL}$, hence all atomic processes can be computed "as is", without bothering which speed of light ($c$ or $\hat{c}$) to use. In that limit, the spatially average radiation is also computed correctly,
\begin{equation}
    \langle N_A\rangle \approx \frac{\langle\dot{N}\rangle}{c\bar\kappa} = \frac{\lambda_{\rm MFP}}{c}\langle\dot{N}\rangle.
    \label{eq:rslthick}
\end{equation}
\item[Cons] In the optically thin limit, photons are not conserved:
\begin{equation}
    N_A(t_1) - N_A(t_0) = \int_{t_0}^{t_1} \frac{\hat{c}}{c}\dot{N} dt \ll \int_{t_0}^{t_1} \dot{N} dt.
    \label{eq:rslthin}
\end{equation}
\end{description}
\end{leftbar}

\smallskip
{\parindent=0pt \bf Flavor B:} 
\begin{leftbar}
\begin{equation}%
\label{eq:momb}%
    \frac{\partial N_B}{\hat{c}\partial t} + \frac{\partial M_B^j}{\partial x^j} = 
    - \kappa N_B + \frac{\dot{N}}{\hat{c}}.
\end{equation}
\begin{description}
\item[Pros] In the optically thin limit, the photons appear to be conserved:
\begin{equation}
    N_B(t_1) - N_B(t_0) = \int_{t_0}^{t_1} \langle\dot{N}\rangle dt.
    \label{eq:bndot}
\end{equation}  
\item[Cons] In the steady-state 
\[
    N_B = \frac{c}{\hat{c}}N_{NL}.
\]
Hence, one has to use a reduced speed of light for some of the atomic processes. For example, the photoionization rate is now
\[
    \Gamma_j = \hat{c} \int \sigma_j(\nu) N_B(\nu) d\nu. 
\]
\end{description}
\end{leftbar}

The latter is, however, \emph{a problem}. In flavor B, one "photon" does not ionize one atom; instead, it ionizes only a $\hat{c}/c$ fraction of an atom. In other words, the photon conservation 
in Equation (\ref{eq:bndot}) is not real - a unit of $N_B$ is not one photon but only a $\hat{c}/c$ fraction of a photon. More than that, if we introduce
\[
\hat{N}_B = \frac{\hat{c}}{c} N_B,
\]
then $\hat{N}_B = N_A$. So, a simple rescaling transforms flavor B into flavor A. Hence, there is in reality just one flavor of the RSL approximation; the difference in formulations found in the literature is merely a choice of notation. And the RSL approximation does undercount ionizations (and hence photons) in the optically thin limit \citep{Cain2024}:
\begin{equation}
    \frac{d}{dt} \langle\Gamma\rangle = 
    \hat{c}\int\sigma_\nu\langle\dot{N}_\nu\rangle d\nu \ll c\int\sigma_\nu\langle\dot{N}_\nu\rangle d\nu
    \label{eq:gammabar}
\end{equation}
(the last term is the correct value).

\section{Counting Missing Photons}

A natural question to ask is whether it is possible to account for the missing photons in the RSL approximation. All terms in Equation (\ref{eq:gammabar}) are known, so there is no reason why it cannot be done. In fact, \citet{Cain2024} presented one approach to how it can be accomplished under some assumptions. 

\subsection{Numerical Scheme}

To demonstrate how it can be done in practice, I use a new simulation framework called ALTAIR (which stands for "Adaptive, Locally Timestepping, Asynchronous Implementation of Refinement"). It is still currently in development and is not ready for public release, and hence will be described in detail elsewhere. However, for the tests presented here, only two features of ALTAIR are important: it implements the atomic and cooling rates from CROC simulations \citep{Gnedin2014}, and it implements several moments-based solvers for radiative transfer, including several flavors of the OTVET approximation \citep{Gnedin2001,Gnedin2014} and the M1 closure approximation as implemented in the RAMSES code \citep{Rosdahl2013}, as well as their generalizations for semi-implicit numerical schemes. A few more details about the framework and examples of small simulations of reionization with ALTAIR are presented in \citet{Gnedin2025}.

Because the details of the numerical implementation are important for the following, I list them here explicitly. All radiative transfer schemes used here are variants of the Variable Eddington Tensor (VET) class of models. Moments equations (\ref{eq:moms}) include 3 unknown quantities ($N$, $M^i$, and $P^{ij}$) in two equations, so a closure relation is needed. This is achieved by adopting an \emph{ansatz} for the Eddington tensor 
\[
    h^{ij} \equiv \frac{P^{ij}}{N}.
\]
Moments equations in the RSL approximation are discretized on a (locally) uniform grid $\vec{x}_{ijk} = (i\Delta x,j\Delta x,k\Delta x)$ to the second order in space and to the first order in time:
\begin{widetext}
\begin{equation}
    \frac{N_{ijk}(t+\Delta t)-N_{ijk}(t)}{\hat{c}\Delta t} + \frac{M^1_{i+1,jk}-M^1_{i-1,jk}+...}{2\Delta x} = 
    -\kappa_{ijk} \tilde N_{ijk} + \frac{\dot{N}_{ijk}}{c} 
    + \frac{1}{2\Delta x}\left(N_{i+1,jk}+N_{i-1,jk}+...-6\tilde N_{ijk}\right),
\end{equation}
\end{widetext}
where dots hide terms with analogous derivatives in $y$ and $z$ directions, and the last term is the global Lax-Friedrich artificial diffusion, without which the scheme is numerically unstable. All terms without the explicit time dependence are evaluated at the initial time $t$. The equation for the first moment $M^i$ is directly analogous.

This equation is stiff, since $\kappa_{ijk}$ can be large. One can design a sequence of semi-implicit schemes by allowing $\tilde N_{ijk}$ to depend on the "new" value of $N_{ijk}(t_1)$,
\[
    \tilde N_{ijk} = \gamma N_{ijk}(t+\Delta t) + (1-\gamma) N_{ijk}(t).
\]
This is equivalent to only using the diagonal part of the Jacobian in a fully implicit numerical scheme.

With this substitution, the "new" value of $N_{ijk}$ is computed explicitly:
\begin{equation}
    N_{ijk}(t+\Delta t) = N_{ijk}(t) + \frac{\alpha}{1+\alpha\gamma(\Delta x\kappa_{ijk}+3)}\mathcal{R},
    \label{eq:num}
\end{equation}
where
\begin{widetext}
\begin{equation}
    \mathcal{R} = -\frac{1}{2}\left(M^1_{i+1,jk}-M^1_{i-1,jk}+...\right) 
    -\Delta x\kappa_{ijk} N_{ijk} + \frac{\Delta x}{c}\dot{N}_{ijk} 
    +\frac{1}{2}\left(N_{i+1,jk}+N_{i-1,jk}+...-6N_{ijk}\right)
\end{equation}
\end{widetext}
is evaluated at the initial time $t$,
\[
    \alpha \equiv \frac{\hat{c}\Delta t}{\Delta x} \leq C,
\]
and $C$ is the Courant-Friedrichs-Lewy (CFL) number for the numerical scheme. The "new" value of the first moment $M^1_{ijk}(t+\Delta t)$ is computed analogously.

In this paper, I will consider 3 different schemes from the full sequence: 
\begin{description}
    \item[explicit]: $C=0.25$, $\gamma=0$ (the original \citet{Rosdahl2013} scheme);
    \item[half-step semi-implicit]: $C=0.5$, $\gamma=0.5$;
    \item[full-step semi-implicit]: $C=0.75$, $\gamma=1$.
\end{description}
All three schemes are robust and numerically stable, and semi-implicit schemes allow larger time-steps because of the larger CFL number $C$.

\subsection{Test Set-up}

It is instructive to start with a simple test: a uniform-density box containing a number of randomly distributed identical ionizing sources. In order to test both a static density field and a cosmological box, I adopt cosmological initial conditions with a 10 cMpc simulation box, $\Omega_M=1$ and $\Omega_B=10^{-4}$ - the low value for the latter allows for running quick tests with modest numerical resolution. For a static density field case, the cosmological expansion is frozen at the value of the scale factor $a=0.1$. The gas is 100\% hydrogen. Radiation sources emit monochromatic 13.6 eV radiation at the rate of $10^{-3}/N_S$ photons per baryon per Myr, where $N_S$ is the number of sources, so that the whole box is ionized in 1 Gyr. To avoid the loss of ionizing photons in recombinations, the recombination rate coefficient is set to a very low number ($10^{-20}\,{\rm cm}^3/{\rm s}$ independent of temperature).

For numerical integration, I use 10 sources to avoid symmetries - with a single source, both the M1 and OTVET approximations are exact until the box is fully ionized, and that hides potential differences between them. I use a $128^3$ uniform grid so that the tests take tens of minutes to run on a typical workstation with RSL and a few hours with the full speed of light. Even with such a low resolution, a sub-percent level of numerical convergence is achieved due to the very low value of $\Omega_B=10^{-4}$, and that serves as a justification for that non-standard choice. 

\subsection{Choice of the Reduced Speed of Light}
\label {sec:vrsl}

\begin{figure*}[t]
\centering
\includegraphics[width=\columnwidth]{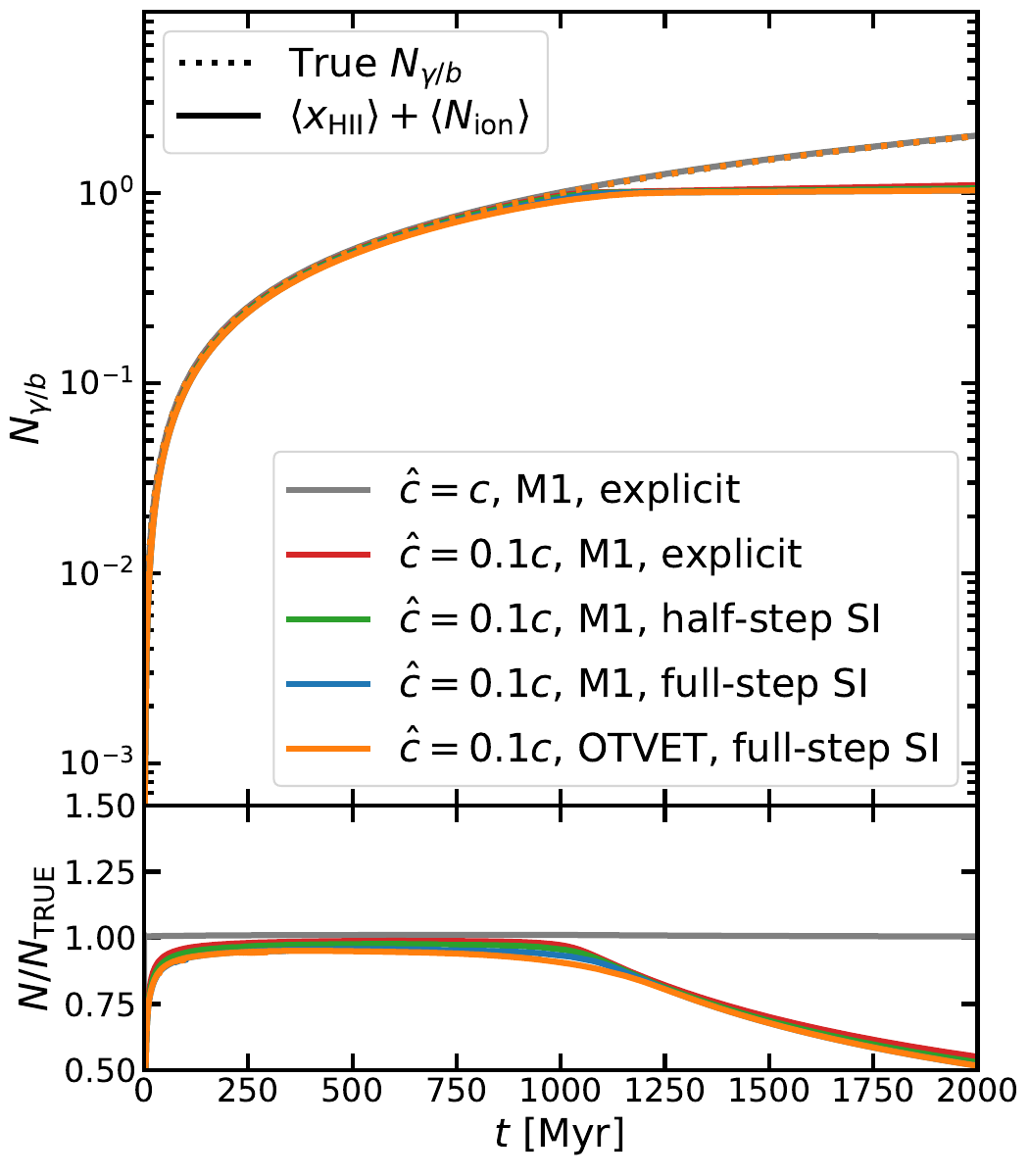}%
\includegraphics[width=\columnwidth]{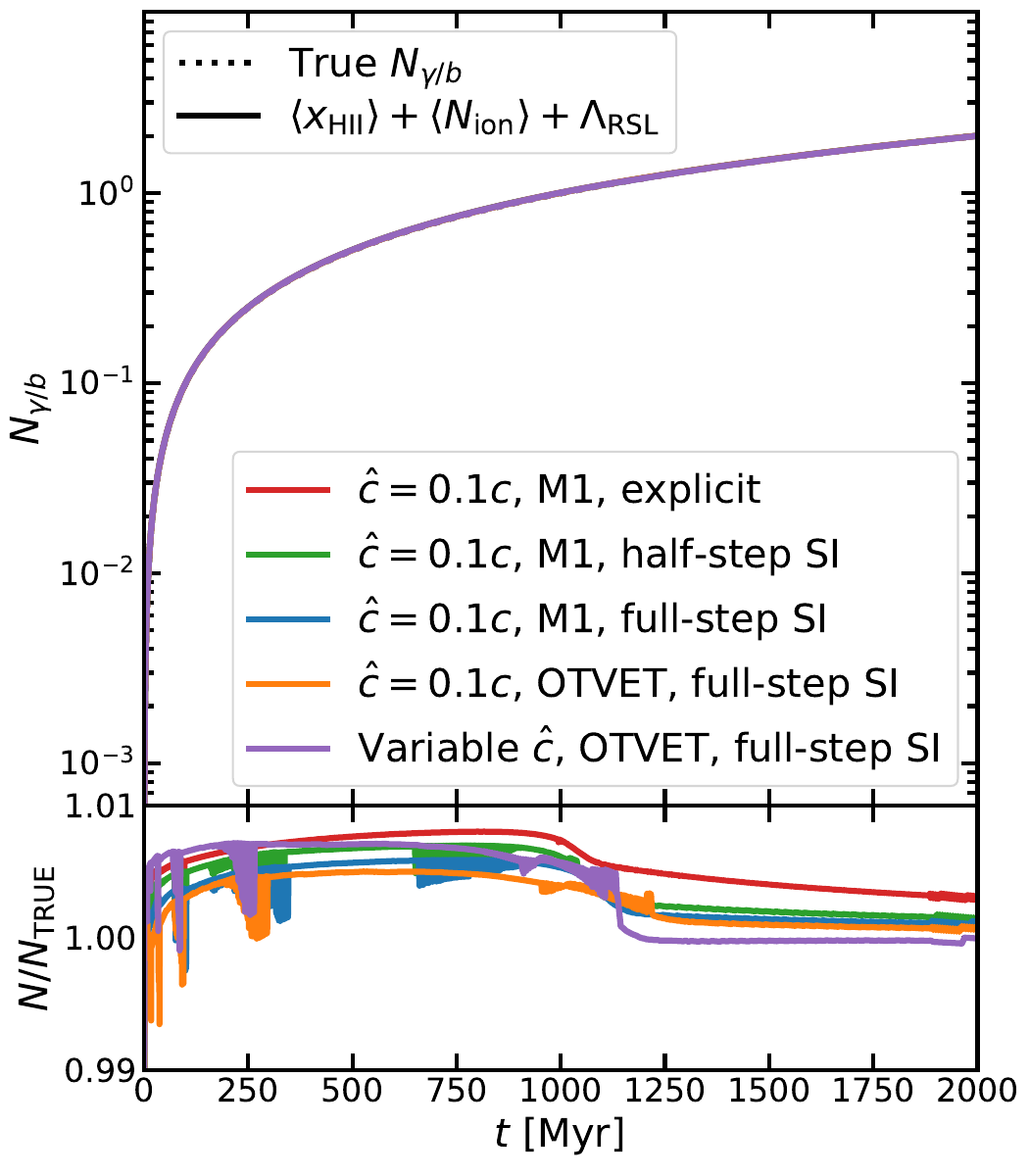}
\caption{The number of ionizing photons per baryon (the main panels) and its ratio to the input value $N_{\rm TRUE} = t/(1\,{\rm Gyr})$ (the basement panels) as a function of time for several numerical schemes. Dotted lines show the true input value, while solid lines show the values counted in the simulation: the sum of all ionizations $\langle x_{\rm HII}\rangle$ and the number of ionizing photons in the radiation field, $N_{\rm ion} = \int N_\nu d\nu$. Ideally, the two are the same. The left panel shows the original schemes from equation (\ref{eq:num}), and the large deviation of the ratio from unity at $t>1\,{\rm Gyr}$ is the manifestation of photon non-conservation from Equations (\ref{eq:rslthin}) and (\ref{eq:gammabar}). The right panel also accounts for the missing photon contribution from Equation (\ref{eq:lambda}). Note that the range of the y-axis in the basement in the right panel is 50 times smaller than in the left panel.} 
\label{fig:nobg}
\end{figure*}

Finally, the choice needs to be made for the value of the reduced speed of light $\hat{c}$. An obvious choice is a fixed fraction of the true speed of light $c$, but that is, in fact, not how the RSL approximation was introduced by \citet{Gnedin2001}. The original formulation did not fix the value of $\hat{c}$ but instead allowed it to vary so that the RT solver made a fixed number of time-steps for each hydrodynamic time-step. If that number is sufficiently large, then the necessary condition for using a Newtonian formulation of the gas dynamics - that all hydrodynamic flows are much slower than the radiation propagation speed - is still satisfied. 

In these tests, the hydrodynamic time-step is actually unconstrained, because with monochromatic sources, there is no photoheating and hence the gas remains isobaric. Thus, and also for the sake of simplicity, I adopt a constant value of $\hat{c} = 0.1c$ as a fiducial choice.  At this value, the reduced speed of light is not too low for the light crossing time to become comparable to other time-scales in the problem (such as the photoionization time), but also is much smaller than $c$, so that all numerical artifacts due to $\hat{c} < c$ are easily identifiable.

In the following, when the variable reduced speed of light is indeed used, the global time-step is set by the requirement that the neutral hydrogen fraction $x_{\rm HI}$ does not change by more than 0.1 in one time-step,
\[
    \Delta t \leq \frac{0.1}{x_{\rm HI}\Gamma},
\]
where $\Gamma$ is the local photoionization rate. The radiative transfer solver takes up to 30 time-steps per global time-step, and that sets the effective reduced speed of light. Occasionally, this set reduced speed of light may be higher than $c$; in that case, the number of radiative transfer solver time-steps is reduced so that $\hat{c}$ never exceeds $c$.

\subsection{Counting Missing Photons}

With the total emission rate of $10^{-3}$ photons per baryon per Myr, the number of emitted ionized photons as a function of time is
\[
    N_{\rm TRUE} = \frac{t}{1\,{\rm Gyr}}.
\]
Since the recombinations are artificially suppressed, the number of photons in the simulation is the sum of the number of hydrogen atoms already ionized, $\langle x_{\rm HII}\rangle$, and the number of ionizing photons still present in the radiation field, $N_{\rm ion} = \int N_\nu d\nu$. Comparison between these two numbers is shown in Figure \ref{fig:nobg} (the left panel) for several variations of the numerical scheme (\ref{eq:num}). 

For a test with the exact speed of light ($\hat{c}=c$, the gray line), the two quantities agree to better than 0.5\%. All the RSL schemes agree with each other (except for some deviations at the start-up, which I ignore): they count photons accurately while the box is being ionized ($t < 1\,{\rm Gyr}$, Equation \ref{eq:rslthick}) and miss most of ionizations after the box becomes optically thin ($t > 1\,{\rm Gyr}$, Equation \ref{eq:rslthin}), exactly as expected.

However, we know everything about the numerical scheme (\ref{eq:num}), and thus the number of photons missed can actually be computed exactly. Equation (\ref{eq:num}) can be rewritten as the increment in the photon number density,
\[
    \Delta N_{\rm RSL} = \frac{\alpha}{1+\alpha\gamma(\Delta x\kappa+3)}\mathcal{R},
\]
where I dropped the grid subscripts $ijk$ for brevity and added a subscript "RSL" to emphasize that this is the photon count in the RSL approximation. Then, the total, true number of photons emitted at the grid location $ijk$ in the time-step $\Delta t$ is
\begin{equation}
    \Delta N_{\rm TOT} = \Delta N_{\rm RSL} \frac{c}{\hat{c}} \left(1+3 \alpha \gamma\right),
    \label{eq:tot}
\end{equation}
and the total number of missing photons in the whole simulation is
\begin{equation}
    \Lambda_{\rm RSL} \equiv \sum\limits_{\rm timesteps} \sum\limits_{ijk} \left(\Delta N_{\rm TOT}-\Delta N_{\rm RSL} \right).
    \label{eq:lambda}
\end{equation}
When this number is added to the sum $\langle x_{\rm HII}\rangle + \langle N_{\rm ion}\rangle$, the photon number conservation is restored to sub-percent precision, as the right panel of Figure \ref{fig:nobg} shows. The same level of precision is achieved when the variable speed of light as described in section \ref{sec:vrsl} is used.
\vspace{1cm}
~

\def\nrsl{N_{\rm RSL}}
\def\ntot{N_{\rm TOT}}
\def\nbg{N_{\rm BG}}
\def\bnbg{\bar{N}_{\rm BG}}

\section{Restoring Missing Photons}

It is awesome that we can count all the missing RSL photons, but unless we can put them back into the simulation, that knowledge remains purely academic. Because the real photons move at the speed of light, and photons in the RSL approximation move at a lower speed $\hat{c}$, there is no general way to figure out where the missing photons are located - figuring this out is equivalent to solving the radiative transfer equation with the full speed of light. Hence, one needs to develop an ansatz for the missing photons.

\citet{Cain2024} proposed one way of restoring the missing photons - adjusting the source emissivity $\dot{N}$ in post-processing. However, they emphasized that this method could not be used to correct the simulation results on the fly and would not be suitable for simulations where the emissivity is also predicted. It is easy to understand why. If $\ntot$ is the exact, full speed of light solution,
\begin{equation}
    \frac{\partial \ntot}{c\partial t} + \frac{\partial M_{\rm TOT}^j }{\partial x^j} = - \kappa\ntot + \frac{\dot{N}}{c},
    \label{eq:ntot}
\end{equation}
then adding the missing RSL photons as an extra source is equivalent to rewriting this equation as
\[
    \frac{\partial \ntot}{\hat{c}\partial t} + \frac{\partial M_{\rm TOT}^j }{\partial x^j} = - \kappa\ntot + \frac{\dot{N}}{c} + \left(\frac{1}{\hat{c}}-\frac{1}{c}\right)\frac{\partial \ntot}{\partial t},
\]
but this equation is, of course, just chicanery. No matter how the last term is evaluated (from a previous time-step, from a predictor step in a predictor-corrector scheme, etc), the numerical solution is going to be violently unstable because the CFL condition is violated by a large factor $c/\hat{c}$. Even if we replace the last term in this equation with its spatial average (i.e., spreading missing RSL photons uniformly in space), numerical instability will persist since all modes, including those with low wavenumbers $k$, are unstable when the CFL condition is violated.

If we want to correct the simulation results as they are running, we need to consider another way of restoring the missing RSL photons. If the reduced speed of light is much lower than $c$ (which is likely to be the case for most applications), then missing photons emitted by a given source are located much further from the source than the counted, "RSL photons". Hence, one can assume that they are less correlated with the source than the RSL photons and count these missing photons as a kind of "background". This is the approach adopted in the CROC simulations \citep{Gnedin2014}.

One can first consider it formally. If $\nrsl$ is the number density of photons in the RSL approximation (equation \ref{eq:moma}):
\[
    \frac{\partial \nrsl}{\hat{c}\partial t} + \frac{\partial M_{\rm RSL}^j }{\partial x^j} = - \kappa\nrsl + \frac{\dot{N}}{c},
\]
then the "background" $\nbg \equiv \ntot - \nrsl$ satisfies the following equation:
\begin{equation}   
    \frac{\partial \nbg}{c\partial t} + \frac{\partial M_{\rm BG}^j }{\partial x^j} = - \kappa\nbg + \left(\frac{1}{\hat{c}}-\frac{1}{c}\right)\frac{\partial \nrsl}{\partial t}.
    \label{eq:nbg}
\end{equation}
The last term in this equation is, in general, spatially variable. However, if one makes another approximation that the source of the background is uniform in space, then the last term in Equation (\ref{eq:nbg}) becomes precisely the rate of emission of the missing RSL photons from Equation (\ref{eq:tot}). However, this equation cannot be used in a simulation, since the $1/c$ coefficient in the first term defeats the whole purpose of using the RSL approximation. One can, however, use a trick from CROC simulations (appendix C from \citet{Gnedin2014}). Namely, Equation (\ref{eq:nbg}) can be used for the mean background only, 
\begin{equation}   
    \frac{d\bnbg}{cdt} = - \bar\kappa\bnbg + \frac{1}{c}\frac{d\Lambda_{\rm RSL}}{dt},
    \label{eq:nbgbar}
\end{equation}
where $\Lambda_{\rm RSL}$ is given by Equation (\ref{eq:lambda}). The time derivative term in this equation has the full speed of light factor in front of it, but since it is an ODE, not a PDE, its computational cost is trivial, so there is no need to use the RSL approximation here.

The full expression for $\nbg$ is then determined by the following ansatz:
\begin{equation}   
    \nbg(\vec{x},t) = \bnbg(t) f_0(\vec{x},t).
    \label{eq:nbgf}
\end{equation}
Substituting (\ref{eq:nbgf}) into (\ref{eq:nbg}) gives the equation for $f_0$,
\begin{equation}   
    \frac{\partial f_0}{c\partial t} + \frac{\partial f_1^j }{\partial x^j} = - \kappa f_0 + \bar\kappa + \frac{1}{c\bnbg}\frac{d\bnbg}{dt}(1-f_0),
    \label{eq:f0}
\end{equation}
with $f_1^j$ being the first moment corresponding to the zeroth moment $f_0$. Now, the background is kind of everywhere, so perhaps it can be modeled in the RSL approximation too. By replacing $c$ in the equation above with $\hat{c}$, I get a complete system of equations that all have the speed of signal propagation equal to $\hat{c}$ and not $c$:
\begin{subequations}%
\label{eq:fmoms}%
\begin{align}
    \frac{\partial f_0}{\hat{c}\partial t} + \frac{\partial f_1^j}{\partial x^j} & = 
    -\left(\kappa+\theta\right) f_0 + \bar\kappa +\theta, \\
    \frac{\partial f_1^i}{\hat{c}\partial t} + \frac{1}{3}\frac{\partial f_0}{\partial x^i} & = 
    - \left(\kappa+\theta\right) f_1^i, 
\end{align}
\end{subequations}
where 
\[
    \theta = \frac{1}{\bnbg}\frac{d\bnbg}{cdt}
\]
and I also assume that the Variable Eddington tensor for the background is isotropic, $h_{\rm BG}^{ij} = \delta^{ij}/3$. 

\begin{figure*}[t]
\centering
\includegraphics[width=\columnwidth]{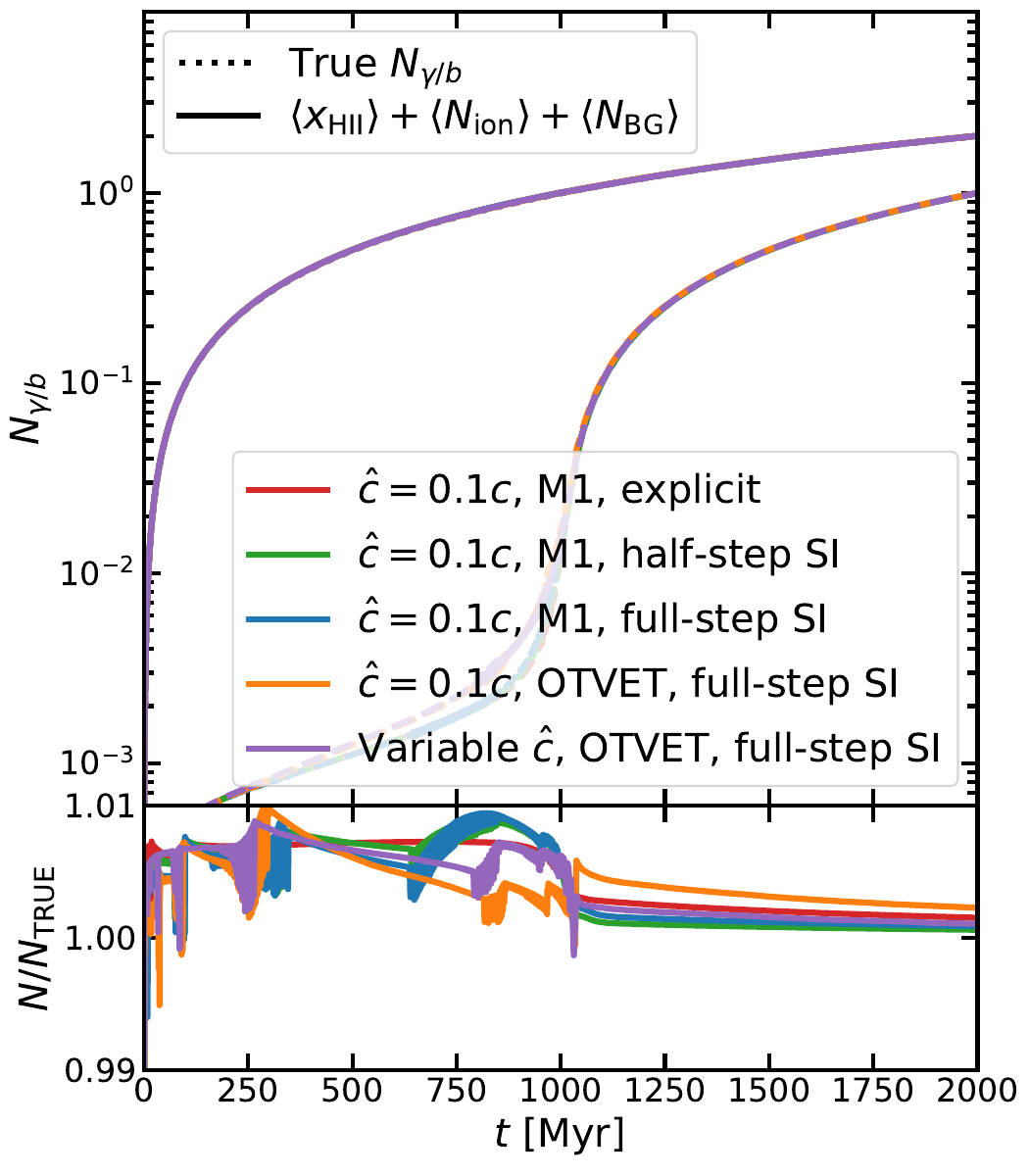}%
\includegraphics[width=\columnwidth]{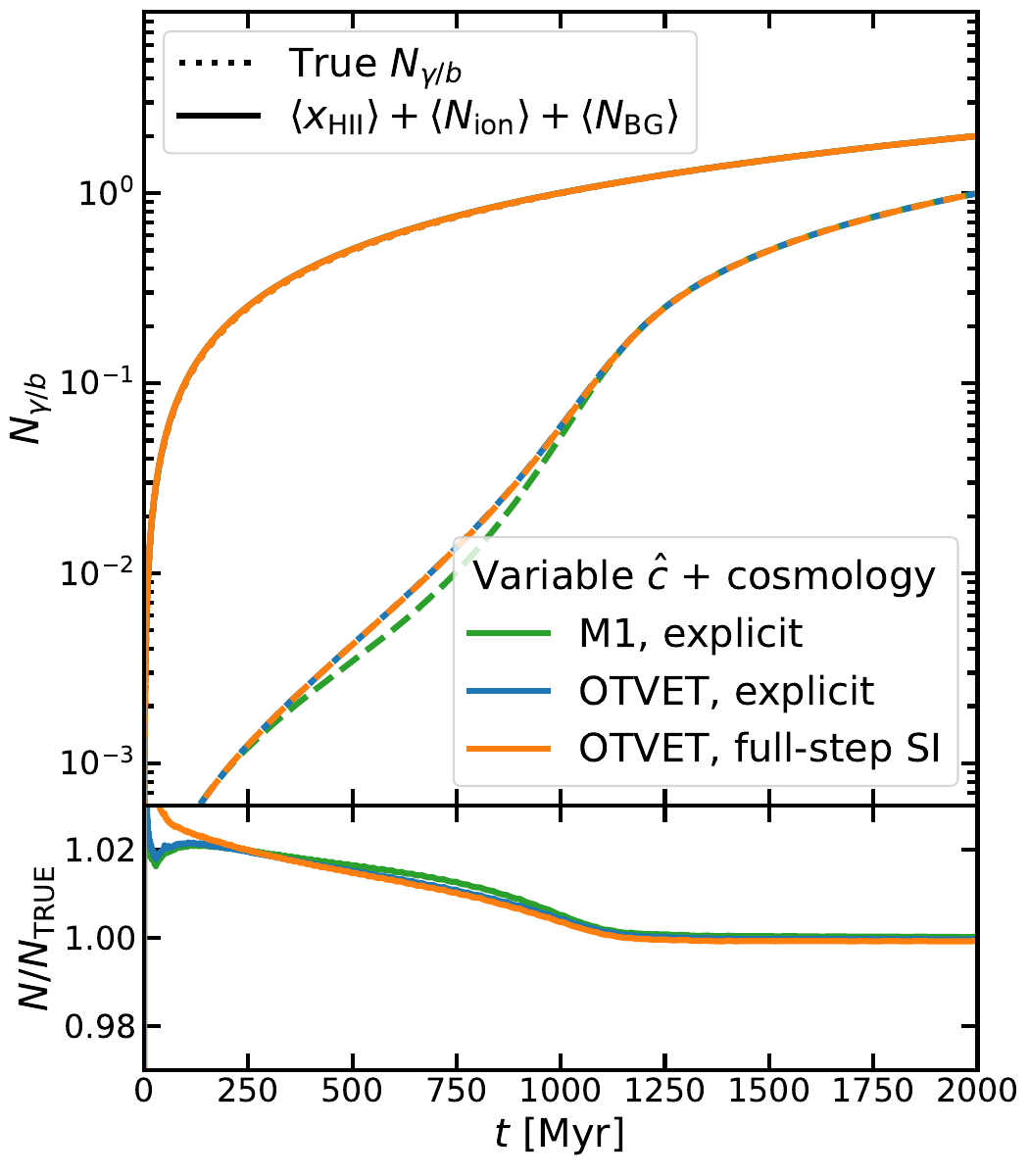}%
\caption{The same as Figure \ref{fig:nobg}, but now with the missing photons included in the background radiation field (left) and several tests with the cosmological expansion included (right).} 
\label{fig:bg}
\end{figure*}

\begin{figure*}[t]
\centering
\includegraphics[width=\textwidth]{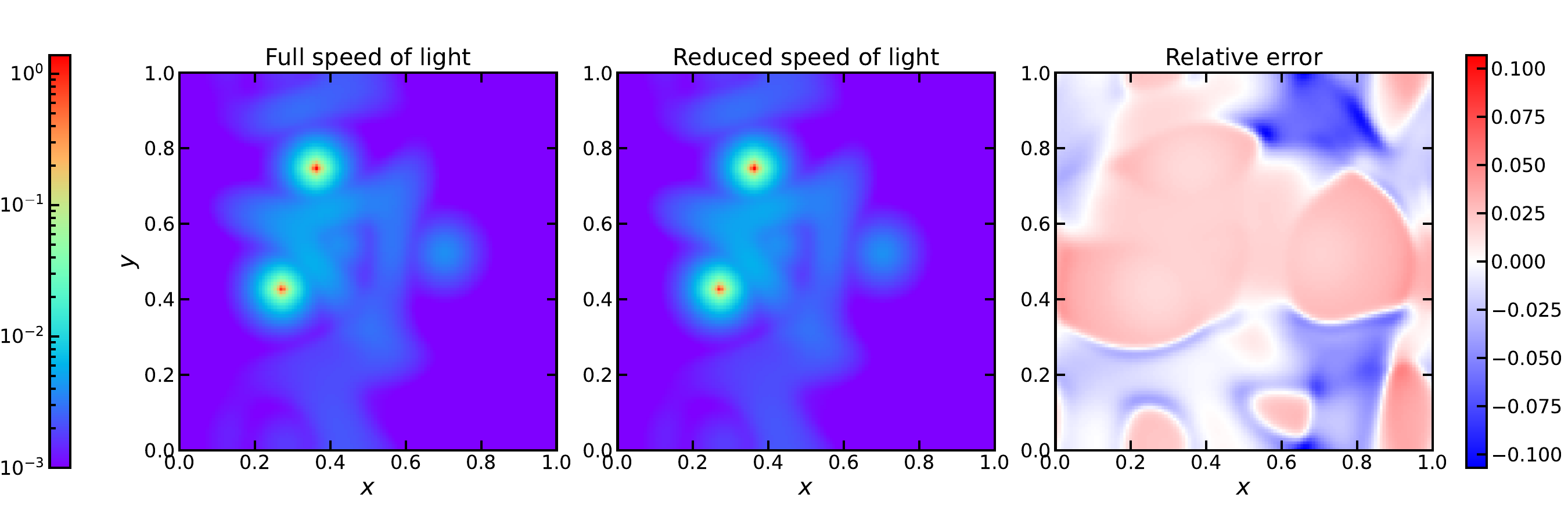}
\caption{Point-wise comparison between the full speed of light solution (left panel) and the $\hat{c}=0.1c$ solution (middle panel) for the explicit M1-closure scheme. The rightmost panel shows the relative difference between the two. The comparison is at $t=500$ Myr, when the errors are approximately the largest.} 
\label{fig:slice}
\end{figure*}

\begin{figure*}[t]
\centering
\includegraphics[width=\columnwidth]{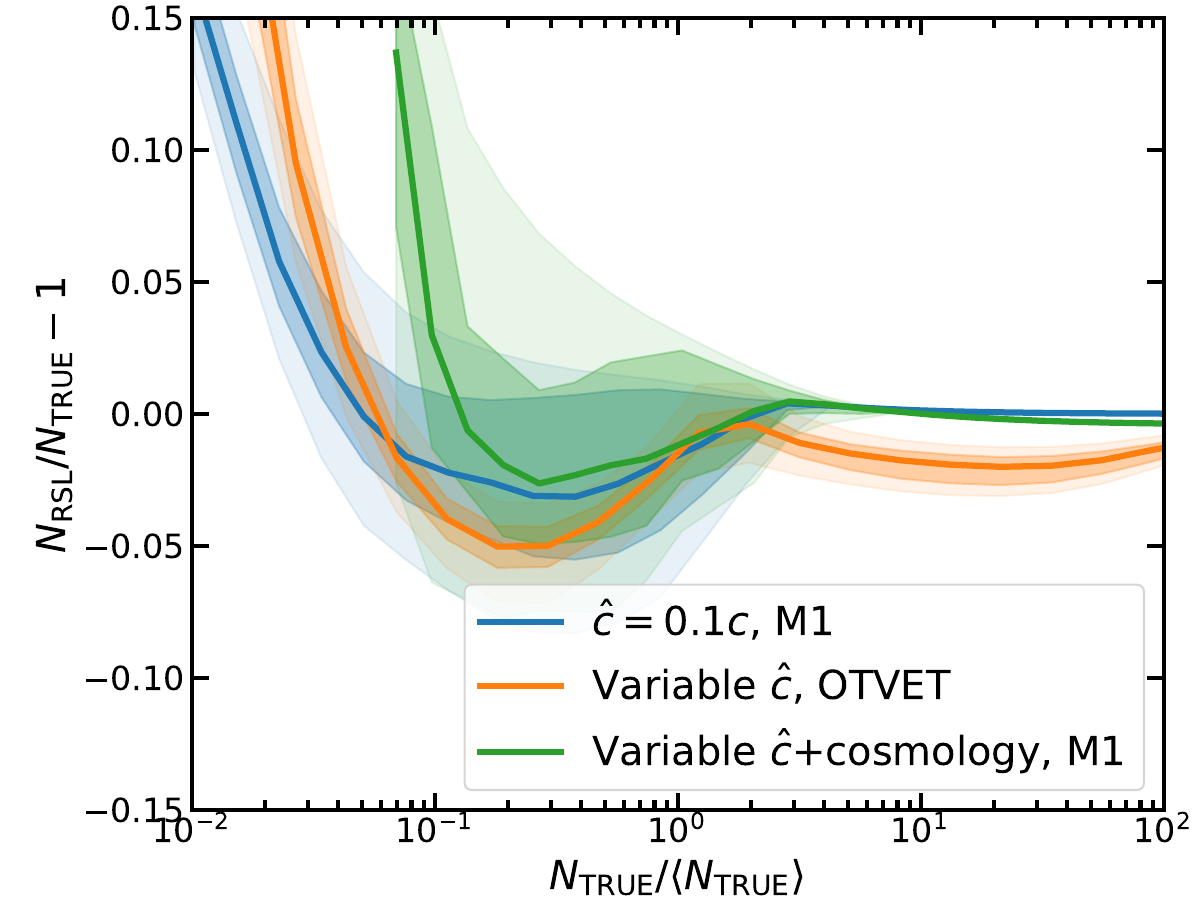}%
\includegraphics[width=\columnwidth]{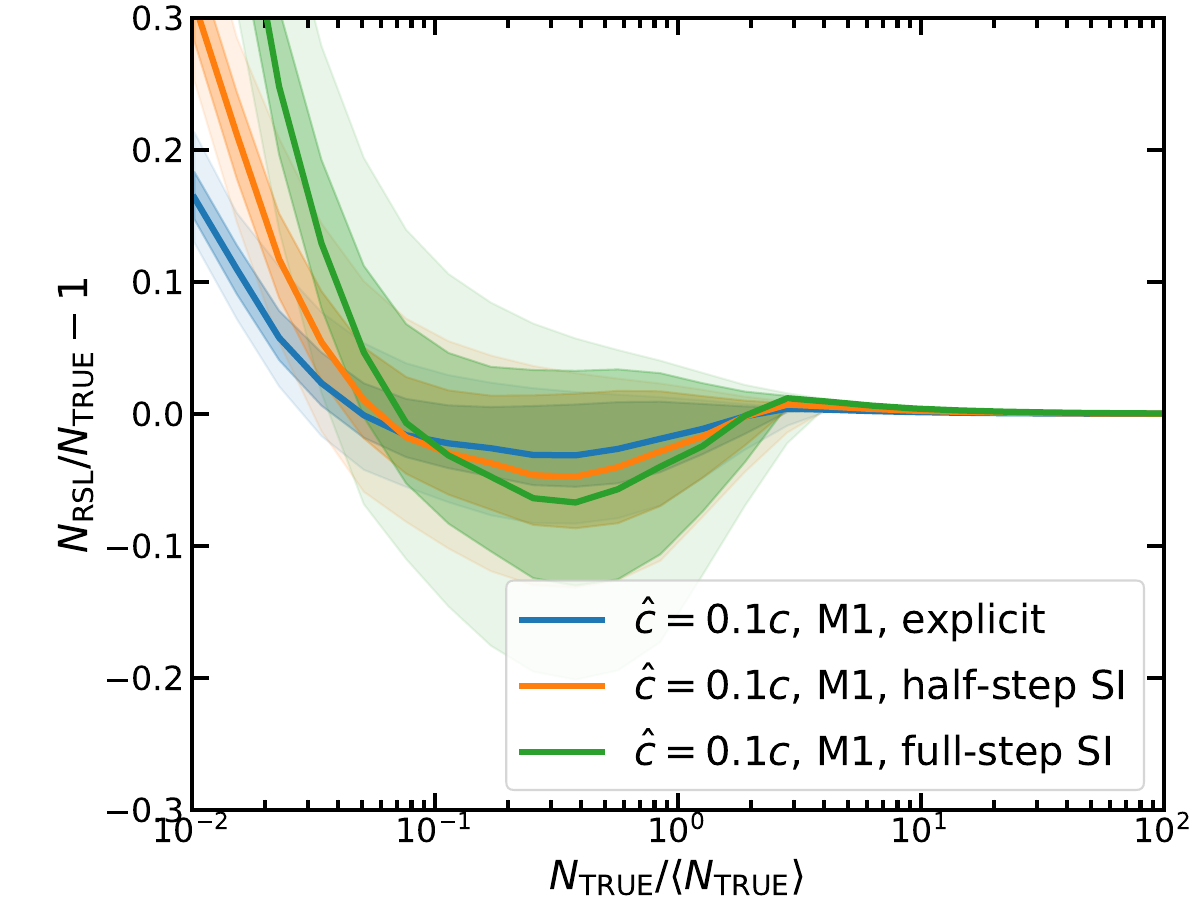}%
\caption{Point-wise error distributions as functions of the "true", full speed of light radiation field for several representative tests, as listed in the legend. Solid lines show the median errors while dark and light shading spans 25\%-75\% and 5\%-95\% percentile ranges, respectively.} 
\label{fig:err}
\end{figure*}

These equations can be solved with the numerical scheme from above - I will call it the "4-equation scheme" hereafter. They can also be further rearranged into a single diffusion-like equation \citep{Gnedin2014}. This scheme will be called the "1-equation scheme". It is more computationally economical and produces results barely distinguishable from the 4-equation scheme for the uniform density tests. Unfortunately, as is shown in the Appendix, it fares noticeably worse for the realistic cosmological tests.

Figure \ref{fig:bg} shows several tests with the missing RSL photons now put into the background component; the second panel in the figure also shows a subset of tests with the cosmological expansion included - with one small twist. With cosmological terms included, equation (\ref{eq:nbgbar}) becomes
\[
    \frac{d\bnbg}{c\,dt} + \frac{H}{c}\left(\nu \frac{\partial \bnbg}{\partial \nu}-3\bnbg\right) = - \bar\kappa\bnbg + \frac{d\Lambda_{\rm RSL}}{c\,dt}.
\]
The first term in the parentheses describes redshifting of photons, and since monochromatic 13.6 eV photons redshift below the hydrogen ionization threshold almost instantaneously, the conservation of ionizing photons is broken. Hence, in these tests, I artificially remove the $\nu {\partial \bnbg}/{\partial \nu}$ term just to be able to check the photon conservation. The precision of the photon number conservation remains at a sub-percent level, except at the very beginning of cosmological tests due to additional interpolation errors in relating cosmic time and the scale factor.

It is not enough to be able to add the missing photons back into the simulation - the added photons also need to be placed in the right locations. Figure \ref{fig:slice} compares the full speed of light solution for the explicit M1 closure (a gray line in the left panel of Figure \ref{fig:nobg}) with the $\hat{c}=0.1c$ case (the red line in the left panel of Figure \ref{fig:bg}). While the average error is sub-percent, the point-wise errors are larger, reaching 10\% in the extreme cases.

Figure \ref{fig:err} quantifies this further by showing the median errors and 25\%-75\% and 5\%-95\% percentile ranges for errors as a function of the "true", full speed of light radiation field. As one can expect, when the radiation field is high, the errors are negligible, but they become larger and biased in the low radiation field regions, where the background component dominates.

\begin{figure*}[t]
\centering
\includegraphics[width=\textwidth]{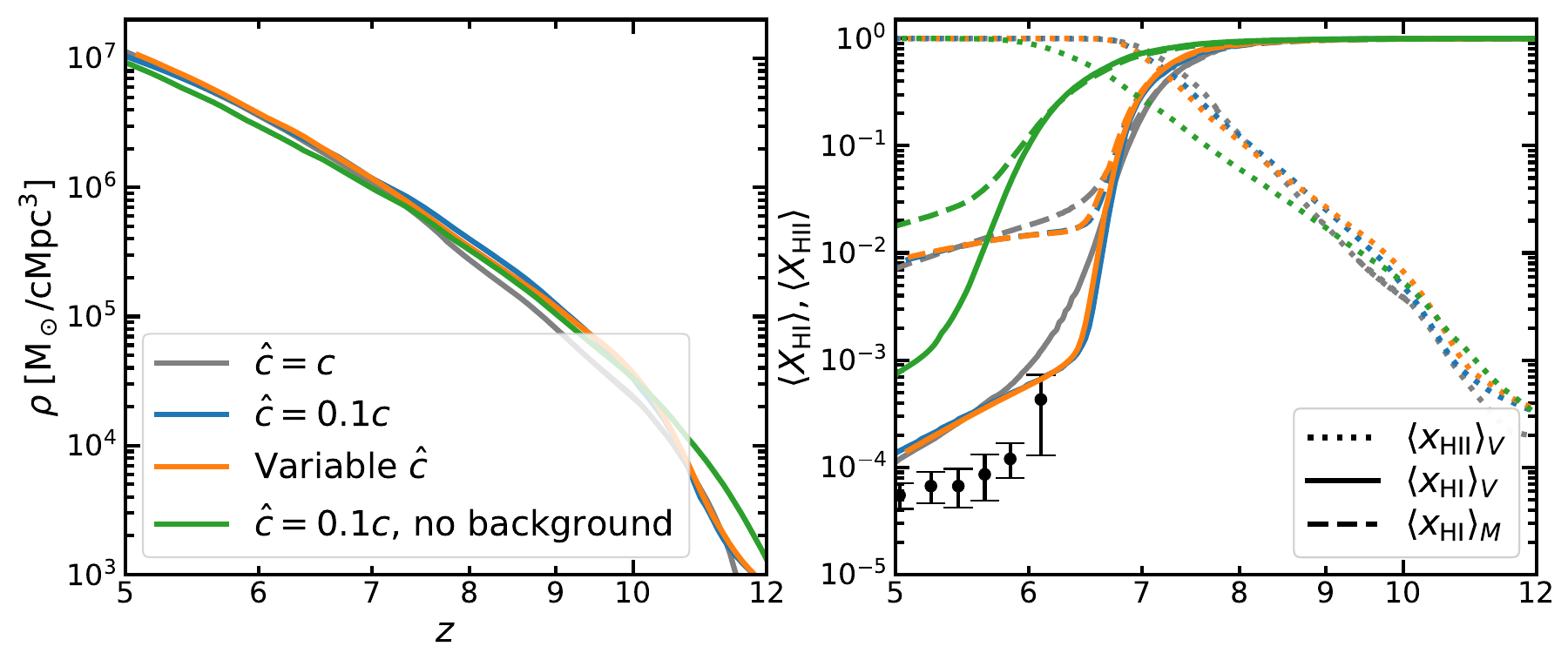}%
\caption{Stellar mass density (left) and hydrogen ionization fraction (right) for four semi-realistic CROC-like simulations of cosmic reionization, as labeled in the legend. The black points with error bars are the data from \citet{Fan2006}. They are shown merely for illustration - the semi-realistic simulations in this figure are not meant to be practical models of reionization.} 
\label{fig:croc}
\end{figure*}

\section{Semi-Realistic Simulations}

Unfortunately, it is very rare that a test can fully verify a numerical method. The ultimate check for how well a fix described above solves the problem can only be done with realistic reionization simulations. With this goal, I perform simulations similar to the  CROC calibration runs: $20h^{-1}$ cMpc box with $256^3$ dark matter particles - this is 8 times smaller than the calibration CROC runs, but since the comparison run must be done with the full speed of light, running it at $512^3$ resolution will require more computational resources than is currently available for this project. The main justification for the original CROC resolution ($512^3$ in a $20h^{-1}$ cMpc box) was to resolve all small-scale structure in the post-reionization IGM. My goal in this paper is to count photons, so mildly under-resolving the low-density IGM can only emphasize the lack of photon conservation - fully resolving all small-scale structure increases the overall absorption. If the total absorption is dominated by the small-scale structure (the Lyman Limit systems), the numerical non-conservation may become hidden.

The maximum refinement level of these semi-realistic simulations is set to ensure that the smallest cell size remains 3.5 comoving kpc (0.5 proper kpc at $z=6$). All other physics in these semi-realistic simulations is designed to mimic CROC simulations, with 2 significant exceptions: the feedback is switched off (to speed up the simulations since it is not important to model galaxies correctly as long as there enough various sources in the simulation volume), and additional absorption due to unresolved Lyman-limit systems is not included (to avoid reducing or hiding discrepancies by absorbing photons in the same way in different tests). 

All simulations were performed with the 4-equation scheme for the best consistency with equations for the local radiation. In the Appendix, I show analogous simulations performed with the 1-equation scheme, which fares much worse. The full speed of light simulation is nontrivially computationally expensive, hence I limit these semi-realistic tests to the M1 closure only (to avoid also running the full speed of light simulation for the OTVET closure).

Figure \ref{fig:croc} shows the global reionization properties of several such semi-realistic simulations: the evolution of the stellar density and ionization history as measured by the volume and mass-weighted neutral fractions. These plots serve as analogs of Figure \ref{fig:bg}. The RSL run with $\hat{c}=0.1c$ without the background has a drastically different reionization history than the full speed of light one. Including the missed photons back into an RSL simulation allows one to reproduce the ionization history of the full speed of light simulation to some extent, although the overlap proceeds faster in the RSL runs. 

\begin{figure}[t]
\centering
\includegraphics[width=\columnwidth]{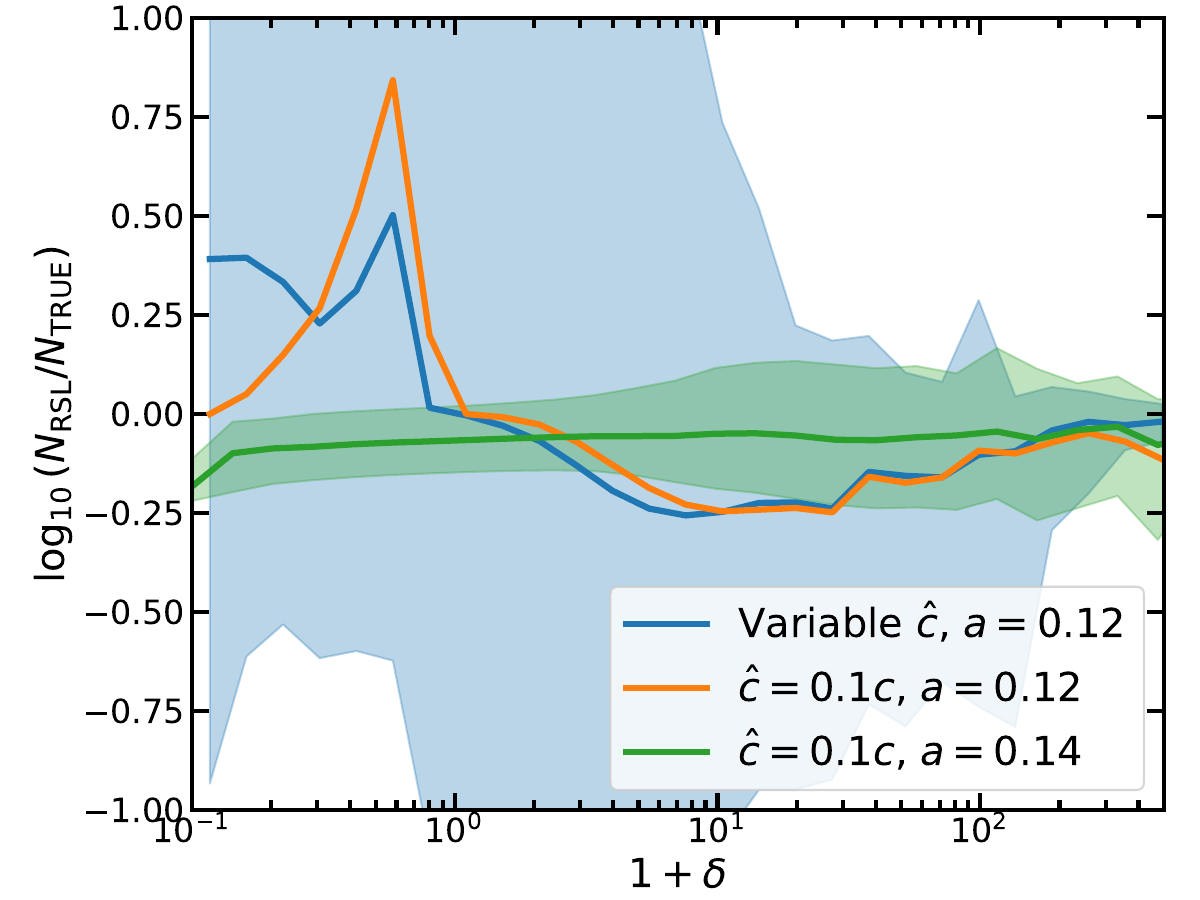}%
\caption{Point-wise error distributions for two semi-realistic simulations (the analog of Figure \ref{fig:err}). The x-axis is now cosmic density, which is a more easily interpretable quantity than the value of the radiation field. Only medians (lines) and 25\% - 75\% quantile ranges (shaded bands) are shown for two runs, to reduce clutter - the errors for the $\hat{c}=0.1c$ run at $a=0.12$ are similar to the variable RSL run at the same epoch.}
\label{fig:cerr}
\end{figure}

Finally, Figure \ref{fig:cerr} shows point-wise error distributions for two semi-realistic simulations at two cosmic times. Unfortunately, for the semi-realistic case, the point-wise errors reach and even exceed an order of magnitude in the voids during the overlap ($a=0.12$). At later times ($a=0.14$), the errors become smaller, but still are of the order of 0.1 dex (25\%) even for the 25\% - 75\% quantile range and higher at higher densities.

\section{Discussion}

The lack of photon conservation is a likely reason for the differences between large reionization simulations like CROC and THESAN \citep{Gnedin2025}. While counting the number of photons missed in an RSL simulation is straightforward (at least for some numerical schemes), adding these photons back is a much taller order. The approach discussed here improves on the ad-hoc scheme used in CROC simulations, but does not completely solve the problem and will need further improvements in the future.

Another potential challenge is the very choice of the "golden solution". While a simulation with the full speed of light may seem to be the one, periodic boundary conditions impose their own artifacts. The background component approach advocated here is largely impervious to these artifacts, so building a proper "golden solution" may, in fact, require larger box, full speed of light simulations, increasing the computational demand for such a solution multifold.

In summary, the overarching goal of this paper is not to offer a solution, but rather to underscore the problem and to encourage the community to think about it. Most likely, only the community-wide effort will pave the way for the next-generation reionization simulations.

\section*{Acknowledgments}

I thank ChatGPT for helping me overcome writer's block and composing the first two paragraphs of the Introduction. This work was supported by Fermi Forward Discovery Group, LLC under Contract No.\ 9243024CSC000002 with the U.S. Department of Energy, Office of Science, Office of High Energy Physics. I also acknowledge the support from the University of Chicago’s Research Computing Center, where the largest simulations used in this work were completed.

\appendix
\section{One-equation Scheme in Semi-Realistic Tests}

Moment equations for the background (\ref{eq:fmoms}) can be rearranged further is one ignores the term $\partial f_1^i/\partial t$. Then
\[
    f_1^i = -\frac{1}{3(\kappa+\theta)}\frac{\partial f_0}{\partial x^i}
\]
and hence
\[
    \frac{\partial f_0}{\hat{c}\partial t} = \frac{1}{3}\frac{\partial}{\partial x^i}\frac{1}{\kappa+\theta}\frac{\partial f_0}{\partial x^i} -\left(\kappa+\theta\right) f_0 + \bar\kappa +\theta.
\]
This diffusion-like equation can be easily solved numerically using a semi-implicit scheme from \citet{Gnedin2014}, Appendix C. This scheme uses 4 times less memory and $\sim$30\% fewer calculations than the default 4-equation scheme.

\begin{figure*}[b]
\centering
\includegraphics[width=0.48\textwidth]{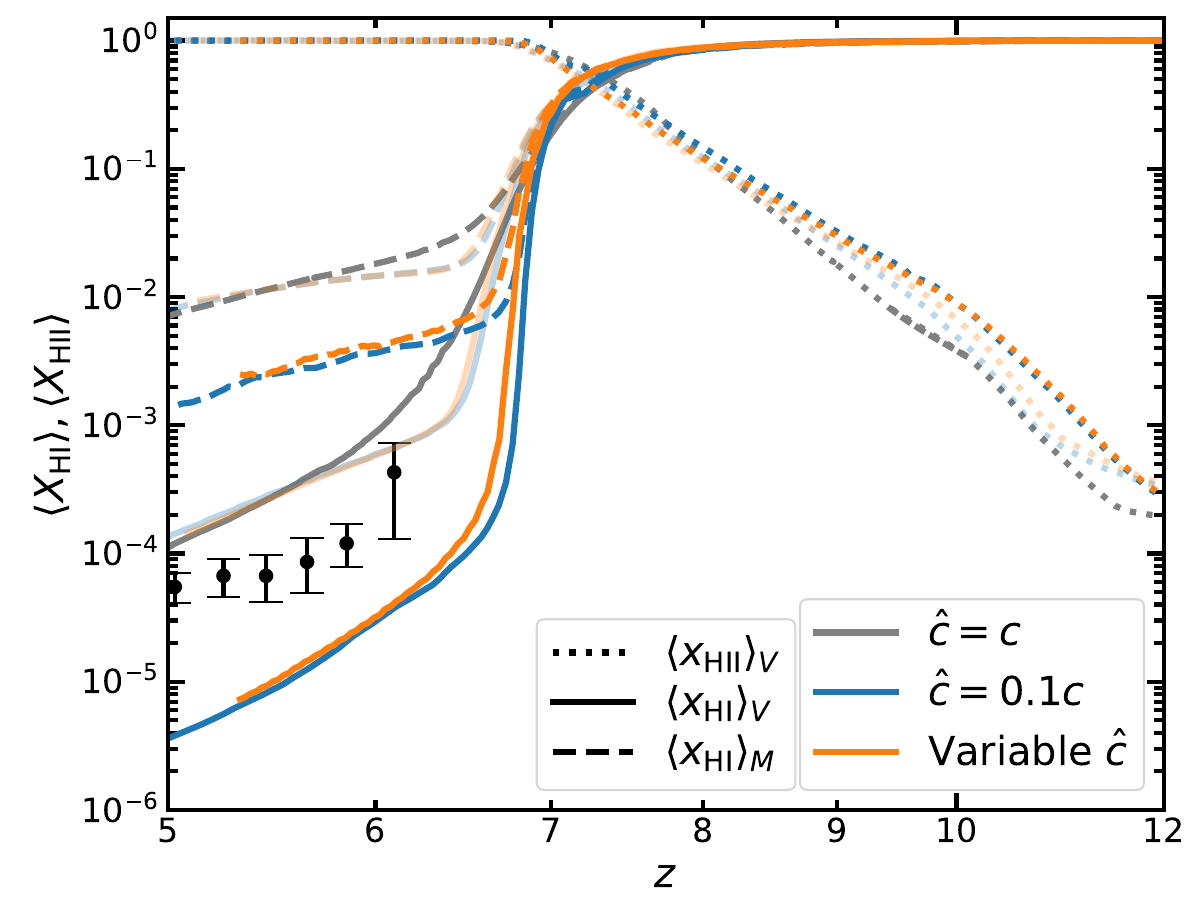}%
\includegraphics[width=0.48\textwidth]{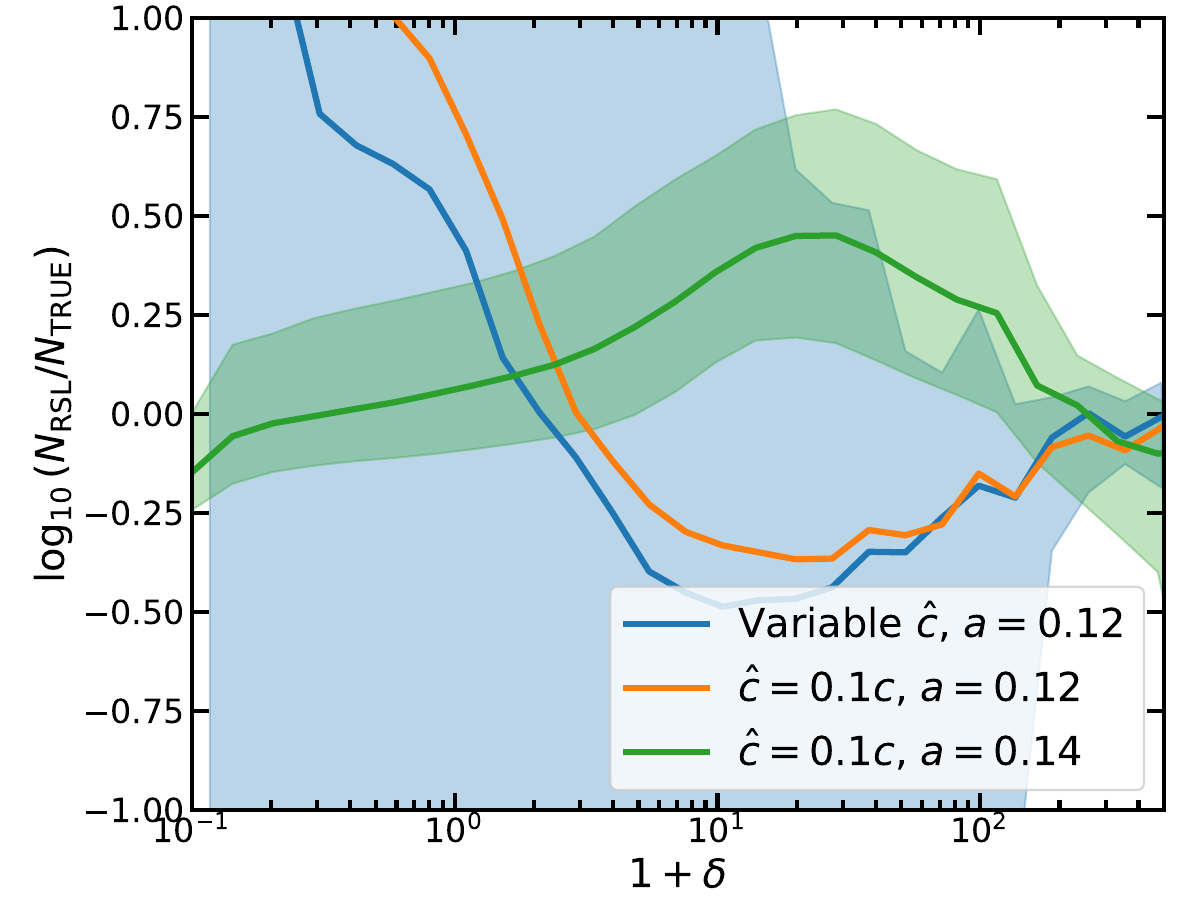}%
\caption{Analogs of the right panel of Figure \ref{fig:cerr} and Figure \ref{fig:cerr} for the 1-equation model. Translucent lines in the right panel show the analogous 4-equation scheme simulation from Figure \ref{fig:cerr} for comparison.} 
\label{fig:1eq}
\end{figure*}

Figure \ref{fig:1eq} shows ionization histories and point-by-point errors for this scheme in semi-realistic tests. Unfortunately, this scheme performs noticeably worse: while the pre-overlap evolution is similar and the overlap occurs at the right time, the post-overlap mean neutral fractions are significantly lower, indicating that the 1-equation scheme puts back missing RSL photons in the wrong places. This is further illustrated by the right panel, showing that the point-wise comparison is worse for the 1-equation scheme than for the 4-equation scheme (Figure \ref{fig:cerr}).

~\newpage

\bibliographystyle{mnras}
\bibliography{main}

\end{document}